\documentclass[12pt,a4paper]{article}  
\usepackage{graphicx}
\usepackage{times}
\def\arcs{$^{\prime\prime}$}                          
\def\farcs{\mbox{\ensuremath{.\!\!^{\prime\prime}}}}
\title{\bf Heated dust around the LMC Wolf-Rayet system \\ 
HD 36402 (BAT99-38)}
\author{P. M. Williams\\
\vspace{1cm}\\
\normalsize Institute for Astronomy, Royal Observatory, Edinburgh, U.K.}
\date{\mbox{}}
\begin{document}
\maketitle
\pagestyle{empty}
%
%
\def\bull{\vrule height .9ex width .8ex depth -.1ex}
\makeatletter
\def\ps@plain{\let\@mkboth\gobbletwo
\def\@oddhead{}\def\@oddfoot{\hfil\tiny\bull\quad
``The multi-wavelength view of hot, massive stars''; 39$^{\rm th}$ Li\`ege Int.\ Astroph.\ Coll., 12-16 July 2010 \quad\bull}%
\def\@evenhead{}\let\@evenfoot\@oddfoot}
\makeatother
%
%
\def\beginrefer{\section*{References}%
\begin{quotation}\mbox{}\par}
\def\refer#1\par{{\setlength{\parindent}{-\leftmargin}\indent#1\par}}
\def\endrefer{\end{quotation}}
%
%
{\noindent\small{\bf Abstract:} 
Infrared photometry of the probable triple WC4 (+O?) +O8I: system HD 36402 (BAT99-38) in the LMC 
shows emission characteristic of heated dust. Although HD 36402 is close to two luminous YSOs, it is 
possible to distinguish its emission at wavelengths less than 10 microns. Simple modelling indicates a 
dust temperature near $800~K$ and mass of about $1.5 \times 10^{-7} M_{\odot}$ amorphous carbon grains. 
The dust emission appears to be variable. It is apparent that Wolf Rayet dust formation occurs also in 
metal-poor environments.}
%
%
\section{Introduction}
Infrared observations show the wide prevalence of dust formation by evolved 
stars of different types: red supergiants, AGB stars, Novae, LBVs, RCB stars -- 
and some Population I Wolf-Rayet stars. Infrared photometric histories, some 
spanning decades, show properties ranging from persistent dust formation at 
apparently constant rates (e.g. WR\,104 = Ve2-45) 
to brief episodes of dust formation at regular intervals (e.g. WR\,140 = HD 193793, 
P = 7.94 y.), and these two stars are considered to be the prototypes of persistent 
and episodic dust makers respectively. 
A handful of systems show intermediate behaviour, forming dust persistently, but 
at a variable rate (e.g. WR\,98a = IRAS 17380-3031). 
The persistent dust makers are nearly all WC9 stars and show a strong preference for 
the metal-rich region of our Galaxy, generally towards the Galactic Centre, while 
the smaller number of episodic dust makers have spectral types ranging from WC4 to 
WC8 and a less concentrated galactic distribution 
-- but all of them are located within the Solar circle.									

In the lower metallicity environments of the Magellanic Clouds, the fraction WR/O stars and 
distribution of WR spectral subtypes are very different from those in our Galaxy. In particular, 
there are no WC stars in the LMC having spectral subtypes `later' than WC4. To date, the LMC WC4 stars									
have not been systematically observed for evidence of dust formation, but comparison of their data 									
in the DENIS (Epchtein et al. 1999) and 2MASS (Skrutskie et al. 2006) surveys suggested that 									
HD~36402 (BAT99-38, Brey 31) varied in $K$ with an amplitude greater than that in $J$ and could be 
a dust maker. This was supported by data in the IRSF Magellanic Clouds survey (Kato et al. 2007). 
The {\em SPITZER} SAGE photometry of massive stars in the LMC (Bonanos et al. 2009) allows us to 
examine the spectral energy distributions (SEDs) of LMC WC4 stars at longer wavelengths, and 
demonstrate that HD 36402 is indeed a dust-making WR star. 
Bonanos et al. tentatively identified HD 36402 with a 70- and 160-$\mu$m 
source but this has to be tested for confusion given the large beam sizes 
(18 and 40 arcsec respectively) at these wavelengths.

\section{Infrared Photometry of HD 36402 from Surveys}

\begin{figure}[h]
\begin{minipage}{10cm}                                                         
\centering
\includegraphics[width=10cm]{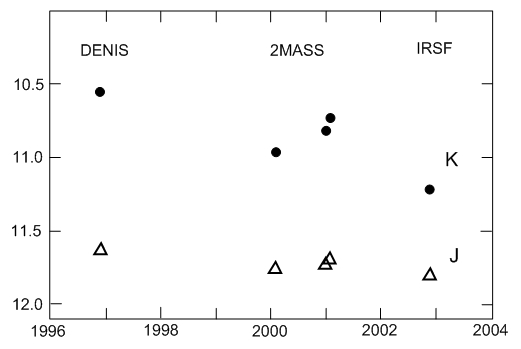}
\caption{$J$ ($\triangle$) and $Ks$ ($\bullet$) magnitudes of HD 36402 from 
the DENIS, 2MASS and IRSF surveys.
\label{LC}}
\end{minipage}
\hfill
\begin{minipage}{6cm}                                                         
\centering
\includegraphics[width=6cm]{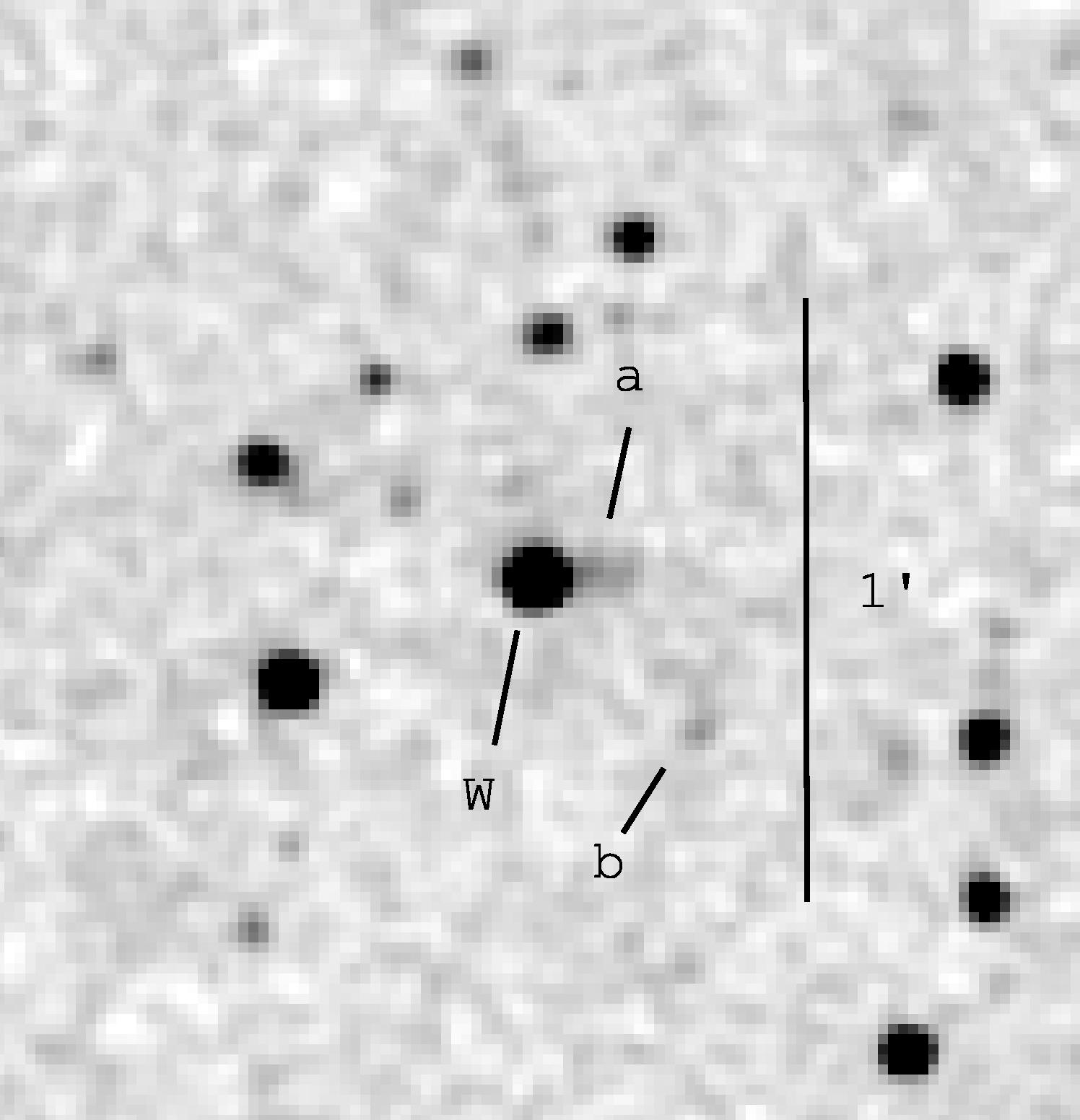}                             
\caption{2MASS $Ks$-band image (N top, E left) of the field of HD 36402 (`W'); 
see text for `a', `b'.}
\label{image}
\end{minipage}
\end{figure}

Photometry and dates of observation of HD 36402 extracted from the DENIS, 
2MASS and IRSF LMC surveys are collected in Table 1 and plotted in Fig.\,\ref{LC}. 
These appear to be only published near-IR photometry of HD 36402 but it is most 
likely to have been observed in other studies. Hopefully, more photometry will 
become available, but the present data are sufficient to show that HD 36402 is 
variable in the near infrared. The amplitude in $Ks$ is significantly greater 
than the uncertainties, $\sigma Ks$. 
It is also greater than that in $J$, arguing against a photospheric origin 
for the variation and suggesting variable emission by hot dust. Given that 
the data come from three different instruments, a $Ks$-band image of the 
field was searched for close companions which might introduce spurious 
variability through confusion. The closest is that marked `a' 
in Fig.\,\ref{image}, 7$^{\prime\prime}$ west of HD 36402. 
This has 2MASS $Ks$ = 14.76, which is $\sim$ 4 mag.\, fainter than the WR star 
and so not a significant contaminant to its near-IR photometry.

\begin{table} [h]
\caption{Near-infrared photometry of HD 36402}
\small
\begin{center} 
\begin{tabular}{lccccc}
\hline 
Survey             & Date    & $J$   & $H$   & $Ks$  & $\sigma Ks$\\
\hline
DENIS strip 4963   & 1996.92 & 11.63 &       & 10.55 & 0.02 \\
2MASS all-sky      & 2000.08 & 11.76 & 11.62 & 10.97 & 0.02 \\
2MASS 6X PSWDB     & 2000.98 & 11.73 & 11.47 & 10.82 & 0.02 \\
2MASS 6X PSWDB     & 2001.09 & 11.69 & 11.46 & 10.73 & 0.02 \\
IRSF LMC0525-6720A & 2002.88 & 11.80 & 11.71 & 11.22 & 0.01 \\
\hline
\end{tabular} 
\end{center} 
\end{table}

\begin{figure}[h]
\centering
\includegraphics[width=9cm]{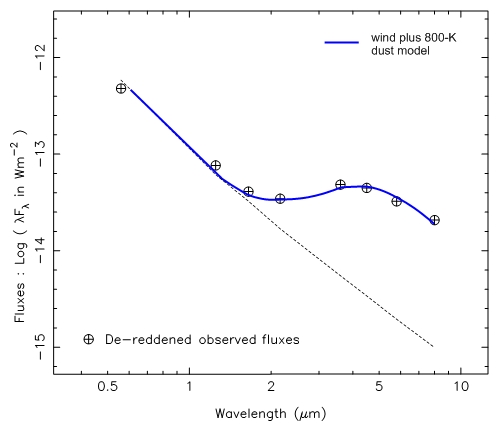}
\caption{De-reddened optical to 8-$\mu$m fluxes of HD 36402 compared with those 
of a model (solid line) comprising a stellar wind (broken line) fitted to the $v$ 
and $J$ data plus emission from an isothermal dust shell.} 
\label{SED}
\end{figure}

The {\em SPITZER} SAGE LMC catalogue gives multi-colour photometry, including 
{\em SPITZER} IRAC (3.6--8.0-$\mu$m) and MIPS (24-$\mu$m) data, for 17 WC4 stars. 
Of these, HD 36402 has the most convincing evidence for a Planckian spectrum, 
peaking near 3.6 $\mu$m, similar to those of dust-making Galactic WC stars (e.g. 
Williams, van der Hucht \& Th\'e 1987, hereafter WHT). The photometry of HD 36402 
was de-reddened using $E_{b-v}$ = 0.09 (Smith, Shara \& Moffat 1990) and the SED 
fitted by a simple stellar wind plus isothermal, optically thin, dust shell model 
following WHT. The dust was assumed to have the emissivity law of amorphous carbon. 
The fit gave $T_d \simeq 800~K$ and $M_d \simeq 1.5 \times 10^{-7} M_{\odot}$, 
adopting a distance of 50 kpc to the LMC. It is illustrated in Fig.\,\ref{SED}.
The dust temperature is somewhat lower than those ($\sim$ 1100~K) found for 
persistent dust makers but may result from the sensitivity of $T_d$ to $Ks$--[3.6] 
and the fact that the $JHKs$ and IRAC 3.6--8.0-$\mu$m data are not contemporaneous.

\begin{figure}[h]
\centering
\includegraphics[width=8cm]{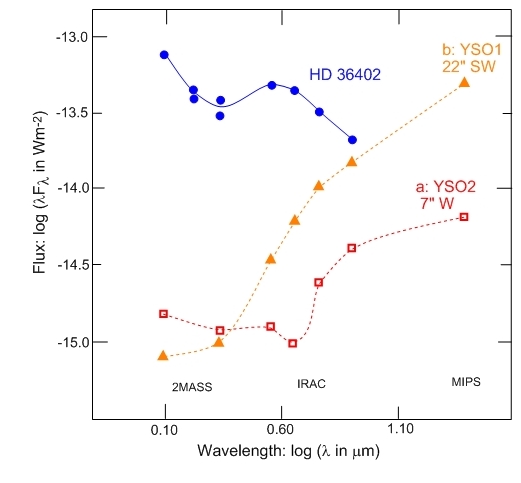}
\caption{Spectral energy distributions of HD 36402 and the nearby YSOs 
discovered by Chu et al.}
\label{YSOs}
\end{figure}

Bonanos et al. tentatively identified HD 36402 with a 70- and 160-$\mu$m source, and 
suggested that its infrared excess might come from its associated ring nebula 
embedded in N51D (Dopita et al. 1994). The environs of HD 36402 are certainly complex. 
Chu et al. (2005) discovered two YSOs close to the WR star. One of them, their YSO-2, 
corresponding to `a' in Fig.\,\ref{image}, is only 7$^{\prime\prime}$ from the WR star. 
The SED of YSO-2 formed from {\em SPITZER} data in Chu et al. and $JHKs$ magnitudes 
from 2MASS is compared with that of the WR star in Fig.\,\ref{YSOs}. At 8 $\mu$m, YSO-2 
is 1.7 mag.\, fainter than the WR star. At this wavelength, the IRAC beam is smaller 
than 2$^{\prime\prime}$ FWHM (Fazio et al. 2004), so we consider YSO-2 well separated 
from HD 36402 in the IRAC data too. At 8 $\mu$m, YSO-1 (`b' in Fig.\,\ref{image}) is 
almost as bright as the WR star (Fig.\,\ref{YSOs}), but is much further 
(22$^{\prime\prime}$) away so the IRAC data can define the SED of HD 36402 in the 
mid-IR without fear of confusion.  

At longer wavelengths, YSO-1 is even brighter (Fig.\,\ref{YSOs} and Chu et al.), and 
its fluxes at 70 and 160 $\mu$m given by Chu et al. are close to those tentatively 
ascribed to the WR star by Bonanos et al. Given the larger beam sizes (18\arcs and 
40\arcs respectively) at these wavelengths, HD 36402 cannot be resolved from the YSOs 
and it is better to ascribe the long-wavelength emission to YSO-1 instead of the WR 
star. Similarly, we do not expect the {\em IRAS} or {\em MSX} observations to resolve 
the WR star from YSO-2.

A more recent observation is the {\em AKARI} IRC (Ishihara et al. 2010) photometry, 
which observed HD 36402 between 2006 May and 2007 August in the 9-$\mu$m band, 
giving S19W = 6.90 (97.9$\pm$6.8 mJy). The larger beam (5\farcs5 at 9 $\mu$m) could 
include some flux from YSO-2, which is rising in this wavelength region. 
From the IRAC and MIPS data, we estimate its flux to be 13.5 mJy at 9 $\mu$m,  
while extrapolation of the HD 36402 SED suggests its 9-$\mu$m flux would be about 
43 mJy (cf. 48 mJy at 8 $\mu$m), three times brighter than YSO-2. 
Hence, the apparent two-fold brightening of HD 36402 at 9 $\mu$m between the 
IRAC observations in mid-2005 and the IRC observation 1--2 years later cannot be 
an artefact of confusion with YSO-2 but must result from changes in the WR dust cloud.  

\section{Conclusion: a WR dust maker in the LMC}

Evidently, HD 36402 is a dust maker like those in the Galaxy. Moffat et al. (1990) deduced 
that it is a triple system: a 3-day binary and more distant O supergiant companion. The inner 
binary separation is probably too small for the O star wind to accelerate to its terminal 
value, and the colliding wind effects probably arise between it and the third component. 
The outer orbit deserves study and relation to the pattern of dust formation, which itself 
requires characterisation from further observations -- some of which may already exist.


%
%
\section*{Acknowledgements}
The author is grateful to the Institute for Astronomy for hospitality 
and continued access to facilities of the Royal Observatory, Edinburgh.
This research has made use of the NASA / IPAC Infrared Science Archive, 
which is operated by the Jet Propulsion Laboratory, California Institute 
of Technology, under contract with the National Aeronautics and Space 
Administration; the Vizier database, operated by the CDS, Strasbourg, 
and the DARTS archive developed and maintained by C-SODA at ISAS/JAXA.
.
%
%
\footnotesize
\beginrefer

\refer Bonanos A.Z., et al., 2009, AJ, 138, 1003

\refer Chu Y.-H., et al., 2005 ApJL, 634, L189

\refer Dopita M. A., Bell J. F., Chu Y.-H., Lozinskaya T. A., 1994, 
         ApJ Supp. 93, 455

\refer Epchtein N., et al. 1999, A\&A, 349, 236

\refer Fazio G. G., et al. 2004, ApJ Supp, 154, 10          

\refer Ishihara, D., et al., 2009, in {\it AKARI, A Light to Illuminate the Misty 
Universe}, eds T. Onaka, G.J. White, T. Nakagawa, I. Yamamura, ASP Conf. Series, 418, 9

\refer Kato D., et al. 2007, PASJ, 59, 615

\refer Moffat A. F. J., Niemela V. S., Marraco H. G., 1990, ApJ, 348, 232

\refer Skrutskie M.F., et al., 2006, AJ, 131, 1163

\refer Smith L. F., Shara, M. M., Moffat A. F. J., 1990, ApJ 348, 471


\refer Williams P.M., van der Hucht, K.A., Th\'e, P.S., 1987, A\&A, 182, 91 (WHT)


\endrefer           
\end{document}